\documentclass[10pt]{iopart}
\usepackage{psfig}
\newcommand{\be}{\begin{equation}}
\newcommand{\ee}{\end{equation}}

\newcommand{\bea}{\begin{eqnarray}}
\newcommand{\eea}{\end{eqnarray}}

\begin{document}
\title[Charmonium in $\gamma$A, pA, and AA]{ 
Charmonium Production in $\gamma$-A, p-A and A-A}

\author{L. Gerland}
\address{School of Physics and Astronomy, Raymond and Beverly 
Sackler, Faculty of Exact Science, Tel Aviv University, Ramat Aviv 
69978, Tel Aviv, Israel}

\begin{abstract}

The data for the production of the $\psi'$ meson in pA collisions at 450 GeV
at CERN-SPS (of the NA50-collaboration)~\cite{na50} yields $\sigma(\psi'
N)\approx 8$~mb under the assumption that the $\psi'$ is produced as a
result of the space-time evolution of a point-like $c\bar c$ pair which
expands with time to the full size of the charmonium state.  However, much
higher values of $\sigma(\psi' N)$ are not ruled out by the data. We show
that recent CERN data confirm the suggestion of ref.~\cite{ger} that color
fluctuations are the major source of suppression of the $J/\psi$ yield as
observed at CERN in both pA and AA collisions. 

\end{abstract}

\section{Introduction}

The NA50-collaboration~\cite{na50} observed quite different $J/\psi$ and
$\psi'$-nucleon cross sections.  Analogous results have been found at
Fermilab~\cite{leitch}: Both experiments found that the charmonium-nucleon
cross section is larger in the target fragmention region than at
midrapidity. This is in good agreement with models that assume that the
$\psi'$ is produced as colorless, point-like $c\bar c$ pair, which expands
with time to its full size and that there exists a relationship between the
spatial distribution of color in a hadron and the cross section of its
interaction with a nucleon. Such a relation is proved in pQCD (perturbative
QCD)~\cite{sigma}. For the nonperturbative regime it is a well known
experimental fact that spatially larger hadrons have larger interaction
cross sections. For example, 3.5 mb for the inelastic $J/\psi$-nucleon cross
section was found at SLAC~\cite{slac} while the inelastic $\pi$-nucleon
cross section at this energies is 20 mb. From charmonium models, e.g. in
ref.~\cite{eich,quigg} it is known that the different charmonium states
($J/\psi$, $\chi$ and $\psi'$) have different spatial sizes. We concluded in
ref.~\cite{ger} that these cross sections are dominated by non-perturbative
contributions.

The agreement of this scenario with the negative $x_F$ Fermilab
data~\cite{leitch} with was first demonstrated in ref.~\cite{ger}. Another
attempt to describe these data with such an expansion scenario was done in
ref.~\cite{arleo}. However, in ref.~\cite{arleo} 20\%-50\% of the
suppression of charmonium states is due to a color octet state added to
the model to explain also the large $x_F$ regime. This state as defined in
ref.~\cite{arleo} is in variance with QCD, because of the following
reason. The eigen life time of this state adjusted to the Fermilab data is
only 0.06~fm. A gluon emitted in such a short time has to have a
momentum of $1/(0.06 {\rm fm})\approx 3.3$~GeV relative to the $c\bar c$
pair~\cite{dok}. In ref.~\cite{arleo} this gluon was assumed to be
massless. 

In ref.~\cite{ger} it was shown that the production of $J/\psi$'s in pA
collisions can be described, if one takes into account the production and
the subsequent decays of higher resonances ($\chi$,$\psi'$) into
$J/\psi$'s.  This leads to a significant increase of the absorption of
$J/\psi$'s as compared to the propagation of genuine $J/\psi$-states. In
ref.~\cite{spieles} it was shown also that the production of $J/\psi$'s in
AA collisions is additionally suppressed by the final state interaction of
charmonium states with newly produced particles like $\pi$'s, $\rho$'s and
so on.

In sect.~\ref{model} the used models are introduced and in
sect.~\ref{data} they are compared to the data. A more detailed
description can be found in ref.~\cite{ger5}. We summarize in
sect.~\ref{conclusion}. 
 
\section{Model Description\label{model}}

\subsection{Semiclassical Glauber Approximation}

The suppression factor $S$ for minimum bias $pA$ collisions can be 
evaluated within the semiclassical approximation (cf.~\cite{yennie}) as
\begin{eqnarray}
S_A&=&\frac{\sigma(pA\rightarrow X)}{A \cdot \sigma(pN\rightarrow X)}\cr 
&=& {1\over A} \int {\rm d}^2B\,{\rm d}z\, \rho(B,z)\cdot \exp \left(-
\int_z^{\infty}\sigma (XN)\rho(B,z'){\rm d}z'\right)\;.
\label{glaub}
\end{eqnarray}
Here $\rho(B,z)$ is the local nuclear ground state density (we used the
standard parametrisation from~\cite{devries}). $\sigma(XN)$ is the
interaction cross section of the charmonium state $X$ with a nucleon and
$\sigma(pA\rightarrow X)$ ($\sigma(pp\rightarrow X)$) is the production 
cross section of the state $X$ in a pA (pp) collision.  We
want to draw attention to the fact that this cross section changes with time
due to the space-time evolution of color fluctuations.  Therefore, it is
necessary to keep $\sigma$ under the integral. 
The suppression factor $S$ of $J/\psi$'s produced in the nuclear medium is
calculated as:
\begin{equation}
S=0.6\cdot ( 0.92\cdot S^{J/\psi}+0.08\cdot S^{\psi'})+0.4\cdot
S^{\chi}\; .
\label{mix}
\end{equation}
Here $S^X$ are the respective suppression factors of the different pure
charmonium states $X$ in nuclear matter.  Eq.~(\ref{mix}) accounts for the
decay of higher resonances after they left the target nucleus into
$J/\psi$'s. The fractions of $J/\psi's$ that are produced in the decays of
higher resonances in eq.~(\ref{mix}) are taken from ref.~\cite{kharzeev}.
However, in ref.~\cite{kharzeev} it is assumed that the different charmonium
states interact with nucleons with the same cross section, which is in
disagreement with the data from the refs.~\cite{na50,leitch}.

In line with the above discussion we want to stress here that
Eqs.~(\ref{glaub})~and~(\ref{mix}) are applicable at CERN energies for
central and negative rapidities, but have to be modified, if applied already
at $y_{c.m.}\sim 0$ at RHIC or higher energies, because at higher energies
charmonium states can be produced outside of the nucleus and the $c\bar c$
pairs propagate through the whole nucleus without forming a hadron. 
Data are often presented in the form $\alpha$ vs. A, where $\alpha$ is 
given by
\begin{equation} 
\sigma_{pA}=\sigma_{pp}\cdot A^\alpha \mbox{ so that } S=A^{\alpha-1}\;.
\label{alpha}
\end{equation}
In the semiclassical Glauber approximation, we take into 
account these color
fluctuations in an effective way as described in ref.~\cite{farrar}. We
assume that charmonium states are produced at $z$ as small $c\bar c$
configurations predominantly through gluon-gluon-fusion, then they evolve --
during the formation time $t_f$ -- to their full size. Please note that
there is up to now no theoretical or experimental proof for the assumption
that charmonium states are produced in point-like configurations as
predicted in pQCD. A way to test this experimentally was suggested recently
in ref.~\cite{ger3}. It's based on the idea that the $\psi'$ and the 
$\psi''$ are mixed states of the $2S$ and the $1D$ charmonium states. The 
names of these states comes from a comparison of nonrelativistic charmonium 
models with the nonrelativistic wave functions of the positronium (actually 
even the name charm{\it{onium}} was given to pronounce the similarity with 
positr{\it{onium}}). Namely \begin{eqnarray}
\left|\psi'\right\rangle&=&\cos \theta \left|2S\right\rangle + \sin\theta
\left|1D\right\rangle\quad,\cr \left|\psi''\right\rangle&=&\cos \theta
\left|1D\right\rangle - \sin\theta \left|2S\right\rangle\quad.
\label{mixing}
\end{eqnarray}
Since only the $S$-wave contributes to the decay of $\psi$ states into
$e^+e^-$-pairs (at least in nonrelativistic charmonium models) the value of
$\theta=19\pm 2^o $ can be determined from the data on the $e^+e^-$ decay
widths of $\psi'$ and $\psi''$. If the production of these charmonium states 
is pointlike, then only the $S$-wave is produced. These leads to an 
universal 
ratio of
\begin{equation}
{\sigma(\psi'')\over\sigma(\psi')}\approx
{\Gamma(\psi''\to l^+l^-)\over \Gamma(\psi'\to l^+l^-)} \approx 0.1\;.
\end{equation}
$\sigma(\psi'')$ resp. $\sigma(\psi')$ are here the production cross 
sections of the $\psi''$ and  resp. $\psi'$ in various processes.  
Predictions for different processes can be found in ref.~\cite{ger3}.

If the formation length of the charmonium
states, $l_f$, becomes larger than the average internucleon distance
($l_f>r_{NN}\approx 1.8$ fm), one has to take into account the evolution of 
the cross sections with the distance from the production 
point~\cite{farrar}. Here we assume motivated by quantum diffusion that the 
cross section increases linearly with time.
The formation length of the $J/\psi$ is given by the energy denominator
$l_f\approx \frac{2p}{m^2_{\psi'}- m^2_{J/\psi}}$, where $p$ is the momentum
of the $J/\psi$ in the rest frame of the target. With $p=30$ GeV, the
momentum of a $J/\psi$ produced at midrapidity at SPS energies
($E_{lab}=200$ AGeV), this yields $l_f\approx 3$ fm, i.e.\ a proper 
formation time of $\tau_f=0.3$ fm. 
As formation time of the $\psi'$ in its rest system we use here the radius
given by nonrelativistic charmonium models, e.g. see the
refs.~\cite{quigg,eich}. This radius is $r=0.45$ fm for the $\psi'$.  
A larger value of
$\tau_f$ for the $\psi'$ is supported also by the extraction of the 
formation
time of the $J/\psi$~\cite{kharzeev2}. Finally the formation time is 
$t_f=\gamma\cdot\tau_f$, where $\gamma$ is the the Lorentz-factor of the 
charmonium state relativ to the nuclear target. For higher gamma factors, 
i.e. at 
higher energies, the formation time becomes larger than the nuclear targets. 
In this regime a hadronic description during the formation time is 
questionable. A partonic model for this energy range was proposed in 
ref.~\cite{ger4}. 

\subsection{Vector Dominance Model for $\gamma$A\label{VDM}}

The VDM (Vector Dominance Model) takes into account only the direct 
diffractive production of the
$J/\psi$ and the $\psi'$, while the GVDM (Generalized Vector Dominance 
Model) accounts also for the non-diagonal
transitions ($\psi'+N\rightarrow J/\psi+N$ and $J/\psi+N\rightarrow
\psi'+N$). The later are needed, because in photoproduction the particles
are produced as point like configurations and develop then to their average
size. In a hadronic model like the GVDM this is taken into account in form
of the interference due to the nondiagonal matrix elements. 
In the GVDM the photoproduction amplitudes $f_{\gamma \psi}$ and $f_{\gamma 
\psi'}$ for the $J/\psi$ and the $\psi'$
are given by~\cite{fs91}
\begin{eqnarray}
f_{\gamma \psi}&=&{e\over f_{\psi}}f_{\psi \psi}+{e\over f_{\psi'}}f_{\psi' 
\psi}\cr 
f_{\gamma \psi'}&=&{e\over f_{\psi}}f_{\psi \psi'}
+{e\over f_{\psi'}} f_{\psi' \psi'}
\label{GVDM}
\end{eqnarray}
Here $f_{\psi}$ and $f_{\psi'}$ are the $J/\psi-\gamma$ and the 
$\psi'-\gamma$ coupling and $f_{VV'}$ are the amplitudes for the processes 
$V+N\rightarrow V'+N$, where $V$ and $V'$ are the $J/\psi$ and the $\psi'$ 
respectively. In the VDM the non-diagonal amplitudes with $V\neq V'$ are 
neglected. The importance of the nondiagonal transitions is evident,because 
the left hand side of eq.~(\ref{GVDM}) is small. If it is neglected as a 
first approximation~\cite{fs91}, then $f_{\psi' \psi}=-{f_{\psi}\over 
f_{\psi'}} f_{\psi \psi}\approx 1.7\cdot f_{\psi \psi}$. And due to the
CPT-theorem $f_{\psi' \psi}=f_{\psi \psi'}$. 

In ref.~\cite{ger2} the GVDM yields approximately $8\pm 2$ mb for the
$\psi'$-nucleon interaction cross section at SPS-energies as can be seen in
Fig.~1. $\omega$ is the laboratory energy of the photon.
The $J/\psi$-nucleon interaction cross section at
SPS-energies approximately 3.5 - 4 mb is used as input into the analysis of
ref.~\cite{ger2}. The accuracy of such GVDM in predicting the $\psi'$N cross
sections is not clear. The above calculation demonstrates 
that implementing
color transparency leads to significantly larger cross sections of the
$\psi'$ N interaction.

\parbox{7.3cm}{
\hspace{-1cm}\psfig{figure=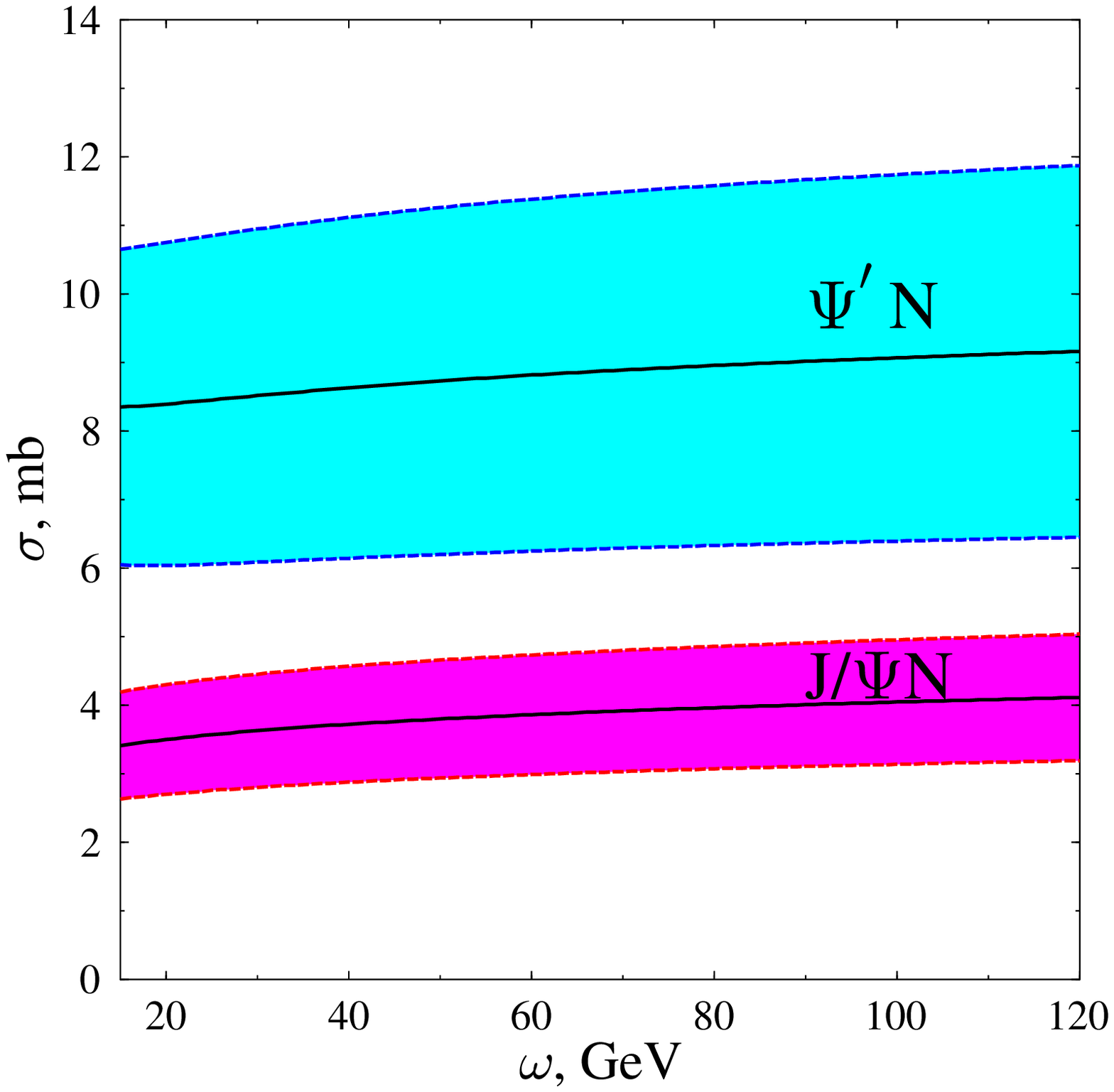,height=8cm}
}
\parbox{5cm}{{\small Figure 1: 
The energy dependence of the elementary charmonium-nucleon
cross sections found in the GVDM. The filled areas show the variation
of the cross sections due to the uncertainty of the experimental $J/\psi$N
cross section.}}
\vspace{-1cm}\\
\section{Comparison with NA50 and NA51 Data\label{data}}

In Fig.~2 we show a comparison between calculations with different
cross sections and different expansion times and the NA50 data~\cite{na50}
for pA collisions and the NA51 data~\cite{na51} for pp and pD collisions for
the cross section of $\psi'$ N interaction vs. the mass of the target. The
y-axis shows $B_{\mu\mu}\sigma_{\psi'}/A$ where $B_{\mu\mu}$ is the
branching ratio for the decay of the $\psi'$ into dimuon pairs. and
$\sigma_{\psi'}$ is the production cross section. The "5.1 mb, instant
formation"  curve in Fig.~2  is the fit of the NA50 collaboration
to their data. Instant formation means that they assumed that the $\psi'$ is
produced with the full cross section and not as a point like particle as in
the description of this paper. (Note the NA50 collaboration fitted
$B_{\mu\mu}\sigma_{\psi'}/\sigma_{DY}$, where $DY$ means Drell-Yan,
we multiplied this fit with the $DY$ cross section in pp
collisions measured by NA51). 

The "8 mb, $t_f=0.45$~fm" curve is the eye-ball fit of the model described
in this paper. For the comparison with the data we need the production cross
section of the $\psi'$ in pp collisions as input. We used here the average
of the pp and pD data of the NA51 collaboration. The value of 8 mb agrees
well with the model parametrisations discussed in the sections~\ref{model}. 
However, we compare also with the calculation with the parameters of
ref.~\cite{ger}, i.e. $\sigma(\psi'N)=20$~mb and $t_f=0.6$~fm. For this
comparison we used the production cross section of the $\psi'$ in pD
collisions divided by two as input. This is also close to the value of the
NA50 fit. One can see in Fig.~2  that the calculation with these
parameters is also in good agreement with the data. A value for
$\sigma(\psi'N)$ of the size of 20 mb is favored by the nucleus-nucleus data
as shown in ref.~\cite{spieles} and in Fig.~3. Plotted is
$B_{\mu\mu}\sigma(J/\psi)/\sigma(DY)$ with the absorption cross sections
from ref.~\cite{ger} and the NA50 data~\cite{na502} for PbPb collisions vs. 
the transverse energy $E_t$, a 
measure for the centrality of the collision. The calculation agrees well 
with the data. The calculation for the $\psi'$ underestimates the data. 
However it is not understood, if this is due to the high value of 
$\sigma(\psi'N)=20$~mb, or if nondiagonal transitions like in 
sect.~\ref{VDM} 
should be taken into 

\parbox{7.3cm}{
\hspace{-1cm}\psfig{figure=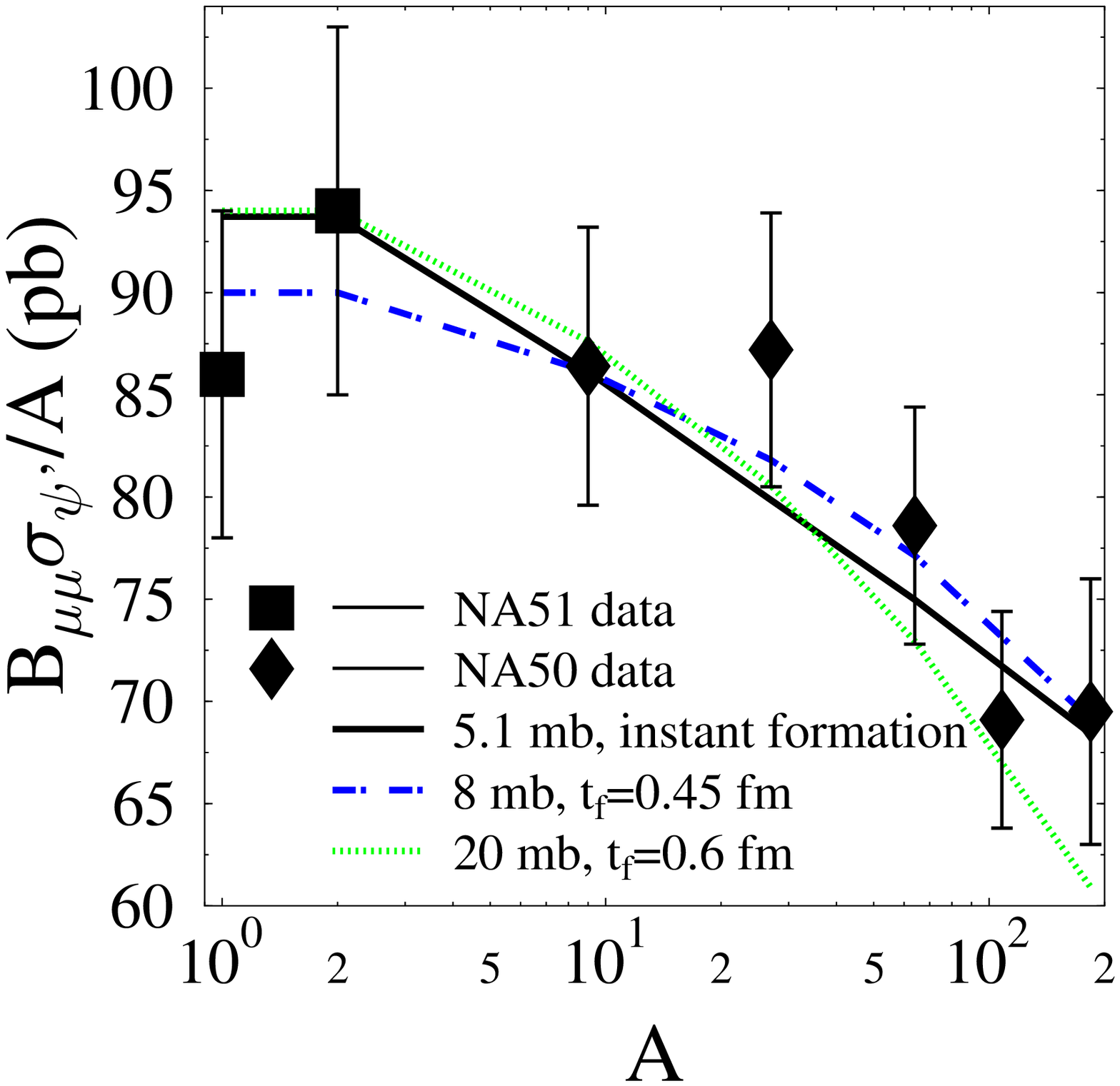,height=8cm}
}
\parbox{5cm}{{\small Figure 2: Plotted are the $B_{\mu\mu}\sigma_{\psi'}/A$
values extracted from calculations with  different absorption cross sections
and different formation times, the NA51 data for pp and pD, and the NA50
data for pBe, pCu, pAg, and pW vs. the mass $A$ of the target.}
}
account in AA collisions, too.

The value of 8 mb is smaller than the theoretical estimate 20 mb of
ref.~\cite{ger}. This is because in ref.~\cite{ger} a formation time of 0.6
fm was chosen for the $\psi'$, while we used her 0.45 fm, the radius of the
$\psi'$ given by the charmonium models. The fact that the formation time is
not known very well is another theoretical uncertainty. Further uncertainty
comes from using diffusion model of expansion at the distances comparable to
the scale of the soft interaction.  Within the error bars the
$\psi'$-nucleon cross section extracted from these pA data and the
prediction of the GVDM, discussed in sect.~\ref{VDM}, are qualitatively
similar. However, further data are needed to learn more about this cross
section. 

\parbox{7.3cm}{
%\hspace{-1cm}
\psfig{figure=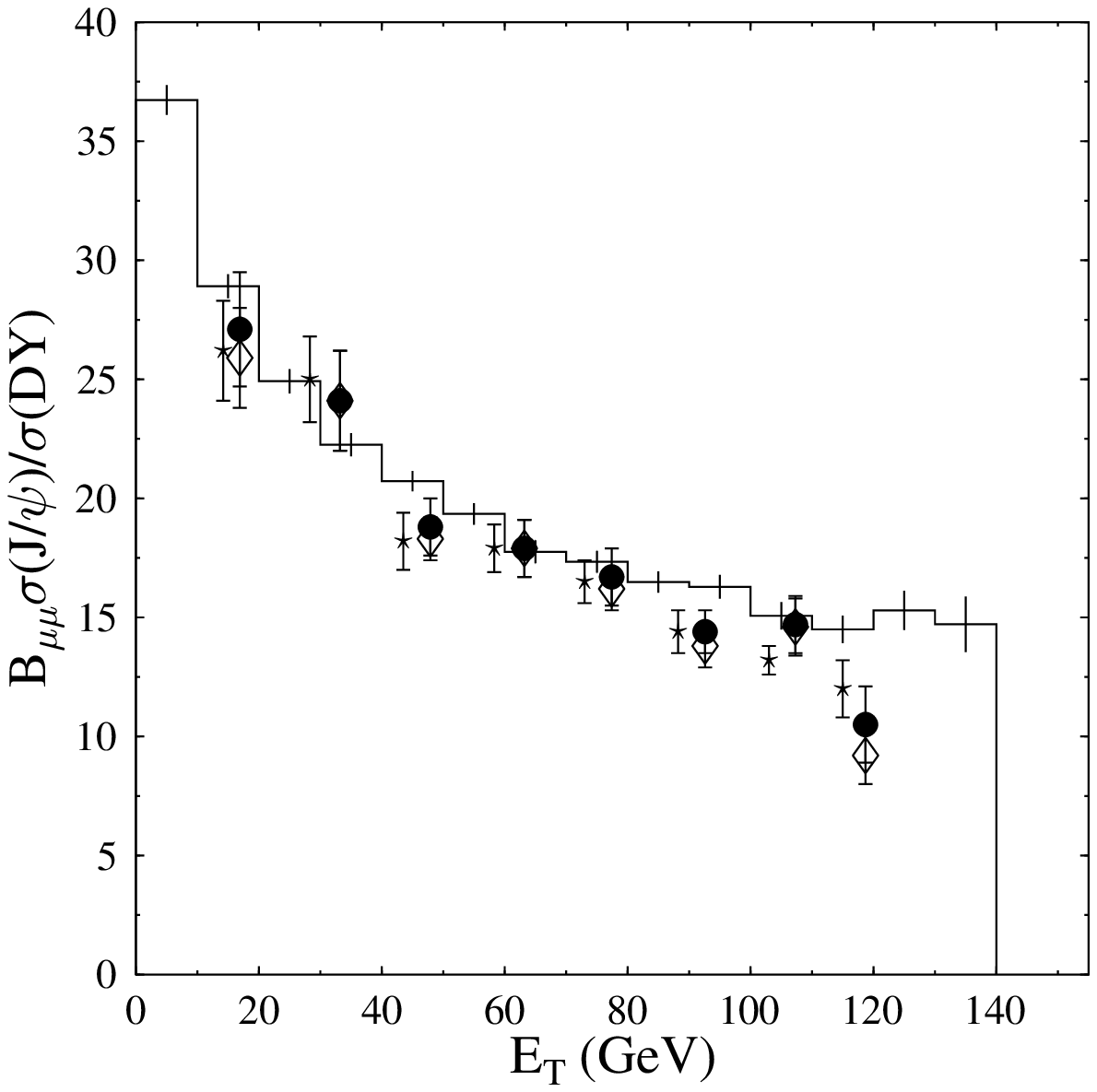,height=8cm}
}
\parbox{5cm}{{\small Figure 3: Plotted is  
$B_{\mu\mu}\sigma(J/\psi)/\sigma(DY)$ with the absorption cross sections 
from ref.~\cite{ger} and the NA50 data for PbPb collisions vs. the 
transverse energy $E_t$}
}

\section{Conclusions\label{conclusion}}

The new data of the NA50-collaboration~\cite{na50} and the data of the
E866-collaboration~\cite{leitch} prove that the $\psi'$-nucleon cross
section is much larger than the $J/\psi$-nucleon cross section.  This is in
agreement with the photoproduction data for these charmonium states as
discussed in the framework of the GVDM in section~\ref{VDM}. This confirms
the QCD prediction that the strength of hadron-hadron interactions depends
on the volume occupied by color. 

Within the assumption that charmonium states are produced as point like
white states, we demonstrated that the data~\cite{na50} can be fitted with a
$\psi'$-nucleon cross section of $\sigma(\psi'N)\approx 8$~mb. However, a 
much larger cross section of e.g.\ $\sigma(\psi'N)\approx 20$~mb is not 
ruled out by the data.  Due to the large experimental errors we conclude 
that the data  and the QCD-motivated models agree, but further data with 
higher accuracy and covering larger rapidity range are needed.

This cross section will be measured soon in proton-nucleus collisions at
HERA B at an energy of $E_{lab}=920$~GeV. The advantage of this experiment
is that it covers a larger range of Feynman-$x_F$, especially in the
negative $x_F$ region. In this region effects due to the formation time of
the hadron will be less important and the genuine cross sections will be
measured. 

\ack
I thank Leonid Frankfurt, Mark Strikman, Horst St\"ocker,
Ramona Vogt, and Mihail Zhalov for discussions, Ramona Vogt for Fig.~3, 
and acknowledge support by the Minerva Foundation and the organizers of the 
SQM2003.

\end{document}